\documentclass[journal]{IEEEtran}
\usepackage{cite}

\ifCLASSINFOpdf
\usepackage[pdftex]{graphicx}
\DeclareGraphicsExtensions{.pdf,.jpeg,.png}
\else
\usepackage[dvips]{graphicx}
\DeclareGraphicsExtensions{.eps}
\fi
\usepackage{epsfig,epsf,epstopdf,graphicx}
\usepackage[cmex10]{amsmath}
\usepackage{amssymb}
\usepackage{amsthm }
\usepackage{nicefrac}
\usepackage{hyperref}
\usepackage{xcolor}
\hypersetup{colorlinks=true}
\usepackage{tabularx}

\theoremstyle{theorem}

\theoremstyle{definition}

\theoremstyle{plain}

\theoremstyle{plain}

\usepackage{algorithm}
\usepackage{multirow}
\usepackage{algcompatible}
\usepackage[bottom]{footmisc}

\PassOptionsToPackage{dvipsnames}{xcolor}
\usepackage{tikz}
\usetikzlibrary{calc}

\ifCLASSOPTIONcompsoc
  \usepackage[caption=false,font=normalsize,labelfont=sf,textfont=sf]{subfig}
\else
  \usepackage[caption=false,font=footnotesize]{subfig}
\fi
\usepackage{fixltx2e}
\usepackage{afterpage}
\usepackage{float}
\usepackage{stfloats}

\makeatletter

\begin{document}

%


\title{A Generalized Framework for Quadratic Noise Modulation Using Non-Gaussian Distributions}

\author{Hadi~Zayyani, \IEEEmembership{Member,~IEEE,} Mohammad Salman, \IEEEmembership{Senior Member,~IEEE},
Felipe A. P. de Figueiredo, and Rausley A. A. de Souza, \IEEEmembership{Senior Member,~IEEE}


\thanks{This work has been partially funded by the xGMobile Project (XGM-AFCCT-2024-9-1-1) with resources from EMBRAPII/MCTI (Grant 052/2023 PPI IoT/Manufatura 4.0), by CNPq (302085/2025-4, 306199/2025-4), and FAPEMIG (APQ-03162-24).}
\thanks{H.~Zayyani is with the Department
of Electrical and Computer Engineering, Qom University of Technology (QUT), Qom, Iran (e-mails: zayyani@qut.ac.ir).}
\thanks{M.~Salman is with College of Engineering and Technology, American University of the Middle East, Egaila, 54200, Kuwait (e-mail: mohammad.salman@aum.edu.kw).}
\thanks{F. A. P. de Figueiredo, and R. A. A. de Souza are with the National Institute of Telecommunications (Inatel), Santa Rita do Sapucaí, Brazil. (e-mail: felipe.figueiredo@inatel.br, rausley@inatel.br).}




\vspace{-1.0cm}}


\maketitle
\thispagestyle{plain}
\pagestyle{plain}

\begin{abstract}
This letter generalizes noise modulation by introducing two voltage biases and employing non-Gaussian noise distributions, such as Mixture of Gaussian (MoG) and Laplacian, in addition to traditional Gaussian noise. The proposed framework doubles the data rate by enabling discrimination in both the mean and variance of transmitted noise symbols. This novel modulation scheme is referred to as Generalized Quadratic Noise Modulation (GQNM). Closed-form expressions for the Bit Error Probability (BEP) are derived for the Generalized Gaussian (GG) and Gaussian Mixture of Two Gaussians (GMoTG) cases. Simulation results demonstrate the advantages of the generalized modulation scheme, particularly under non-Gaussian noise assumptions, highlighting its potential for enhanced performance in low-power and secure communication systems.
\end{abstract}


%
\IEEEpeerreviewmaketitle

\section{Introduction}
\label{sec:Intro}

\IEEEPARstart{N}{oise} modulation is a new modulation paradigm for treating noise as a friend instead of an enemy of communication systems. The idea of noise modulation dates back to Kish's work in 2005 by using two parabolic antennas driven by variable resistances (open, short, and 50 ohms) \cite{Kish05}. Moreover, the thermal noise communication using Kirchhoff-Law-Jonson-Noise (KLJN) is proposed in \cite{Kish06}, \cite{Kish06_1}. In noise communication, which can be wired or wireless, the different variances of the noise generated by using low and high resistors are used to convey the information bits. Some works are done to improve the KLJN noise communication schemes \cite{Ming08}-\cite{Yapici25}. In addition, from a communication engineering perspective, Basar recently formulated a new framework for the BEP calculation of the KLJN noise communication system and suggested two novel detectors that further reduce the BEP \cite{Basar23}. Recently, \cite{Tasci25} proposed a Flip-KLJN secure noise communication where a pre-agreed intermediate level, such as high/low (H/L), triggers a flip of the bit map value during the bit exchange period. More recently, a KLJN-based thermal noise modulation has been proposed, which represents a viable solution for secure Internet of Things (IoT) communication at ultra-low power levels \cite{Salem25}. It proposes a new asymmetric KLJN-based modulation scheme that utilizes a four-resistor structure to improve the bit error rate without requiring additional noise samples per bit. In a recent paper \cite{Basar24}, Basar proposed a noise modulation scheme in which the noise samples with different variances are directly fed to the wireless antenna. In that work, the Bit Error Probability (BEP) is calculated in closed form, and the Noise Modulation system performance is evaluated under fading channels. Also, optimal detection and performance analysis of a thermal noise modulation is investigated in \cite{Alshaw24}. Recently, an on-off digital modulation noise is presented in \cite{Anjos25} and an innovative joint energy harvesting and communication scheme is suggested for future Internet-of-Things (IoT) devices by leveraging the emerging noise modulation technique \cite{Yapici25}.

In this paper, the concept of noise modulation, which is presented in \cite{Basar24}, is generalized under two aspects. Firstly, we add low and high biases to the transmitted noise voltage to have an extra freedom of distinguishability in the mean, giving rise to a novel modulation scheme termed quadratic noise modulation (QNM). It can be considered as a new dimension to the noise modulation scheme to convey another bit in a symbol duration. Therefore, it doubles the data rate of the generalized noise modulation. The similar idea of using non-zero means is also proposed in \cite{Yapici25}, but they use the biases as for energy harvesting in the receiver side not for increasing the dimensionality of the modulation. The second aspect is to use synthetic non-Gaussian random noise samples derived from, for example, a Mixture of Gaussians (MoG) or Laplacian, instead of a Gaussian distribution. As can be seen experimentally in this letter, it improves the sub-bit BEP in the variance dimension. Moreover, in this letter, three study cases of Generalized Gaussian (GG), Generalized Gaussian Mixture of Two Gaussians (GMoTG), and Generalized LAPlacian (GLAP) are discussed, and the closed-form formulas for the BEPs are derived for the cases of GG and GMoTG. The BEP calculation of GLAP is challenging and is left for future work. Simulation results show the advantages of the proposed Generalized QNM (GQNM) approaches.


\section{System Model and Preliminaries}\label{sec:ProblemForm}

In this section, we briefly present the concept of noise modulation as introduced in \cite{Basar24}. A single information bit is used to select a generated random Gaussian noise source, for instance, from a low resistor, $R_L$, or a high resistor, $R_H$. For a bit '0', a low-variance Gaussian noise is transmitted, whereas for a bit '1', a high-variance noise is transmitted. The variance is treated as a distinguishable dimension (or degree of freedom). The noise samples, comprising $N$ samples per bit duration, are fed to a wireless antenna system via a baseband or even an IQ-modulator \cite{Basar24}. At the receiver, the baseband samples are processed for bit detection. This is typically achieved through threshold-based detection by estimating the sample variance to recover the information embedded in the noise variance. This classical noise modulation scheme is analogous to a binary communication system. Next, we propose the GQNM scheme.


\section{Generalized Quadratic Noise Modulation}\label{sec: prop}

In this section, the proposed GQNM scheme is presented. The scheme is shown in Fig.~1. In this scheme, we have two information bits, $b_0$ and $b_1$. The first one, $b_0$, is the bit that selects the voltage bias, $m_L$ or $m_H$. If the bit $b_0=0$, the voltage has a mean of $m_L$, and if the bit $b_0=1$, the mean is $m_H$. The second bit, $b_1$, selects the general distribution of random noise with high or low variance. If $b_1=0$, then the general distribution with low variance, $f_v(v|L)$, is selected, and if $b_1=1$, then the general distribution with high variance, $f_v(v|H)$, is selected. In this study, we use a random noise generator that may have a distribution other than Gaussian, which was used in the classical noise modulation scheme in \cite{Basar24}. The GQNM scheme proposed in this letter doubles the data rate by using the separability of Gaussian or non-Gaussian distributions in both mean and variance dimensions.


Although the proposed scheme is a form of quadratic noise modulation, it can be generalized to higher dimensions. This is achieved by employing noise distributions with more degrees of freedom (i.e., parameters), thereby utilizing more than just the mean and variance as measures of distinguishability. Here, we assume the different distributions for $f_v(v|L)$ and $f_v(v|H)$ are 1) zero-mean, 2) even, and 3) encompass parameters to have distinguishable variances. Examples of such distributions are the Gaussian itself, the Mixture of two zero-mean Gaussians, and the Laplacian. In this case, each symbol duration, which consists of two bits, $(b_0,b_1)$, has four different cases: $00$, $01$, $10$, and $11$. Therefore, the transmitted voltage, $x_n$, might assume four different distributions: $f_v(v-m_L|L)$, $f_v(v-m_L|H)$, $f_v(v-m_H|L)$, and $f_v(v-m_H|H)$, respective to the four two-bit cases. These are the zero-mean general distributions with resulting $m_L$ or $m_H$ means. An example of a waveform generated by the proposed GQNM scheme is shown in Fig.~2.
The assumption that $f_v(v|L)$ and $f_v(v|H)$ are distinguishable in variance means that their underlying parameters can be controlled to yield different values for their variance.
The different distributions $f(x_n|00)=f_v(v-m_L|L)$ and $f(x_n|01)=f_v(v-m_L|H)$ as well as $f(x_n|10)=f_v(v-m_H|L)$ and $f(x_n|11)=f_v(v-m_H|H)$ are distinguishable by
\begin{align}
\label{eq: b1det}
\hat{b}_1=\left\{
            \begin{array}{ll}
              0, & \mathrm{Var}(r_n)<\mathrm{Th}_v, \\
              1, & \mathrm{Var}(r_n)>\mathrm{Th}_v,
            \end{array}
          \right.
\end{align}
where $r_n=x_n+w_n$ is the $n$-th sample of the received signal with $1\le n\le N$, $N$ is the total number of samples in a symbol duration of $T_s$, $w_n$ is the noise term with zero-mean and variance equal to $\sigma^2_w$, and $\mathrm{Th}_v$ is the threshold for distinguishing the variances. The variance threshold, $\mathrm{Th}_v$, can be set as the middle value between low and high values of variances, i.e., $\mathrm{Th}_v=\frac{\mathrm{Var}_{max}+\mathrm{Var}_{min}}{2}$. Moreover, the bit $b_0$ is distinguishable by the different means of the general distributions $f(x_n|00)=f_v(v-m_L|L)$ and $f(x_n|10)=f_v(v-m_H|L)$ or $f(x_n|01)=f_v(v-m_L|H)$ and $f(x_n|11)=f_v(v-m_H|H)$. Thus, in the detector, we have
\begin{align}
\label{eq: b0det}
\hat{b}_0=\left\{
            \begin{array}{ll}
              0, & \mathrm{Mean}(r_n)<\mathrm{Th}_m, \\
              1, & \mathrm{Mean}(r_n)>\mathrm{Th}_m,
            \end{array}
          \right.
\end{align}
where $\mathrm{Th}_m$ is the mean threshold and can be selected simply as the middle value between $m_L$ and $m_H$, i.e., $\mathrm{Th}_m=\frac{m_L+m_H}{2}$.

\begin{figure}
\begin{center}
\includegraphics[scale=0.38]{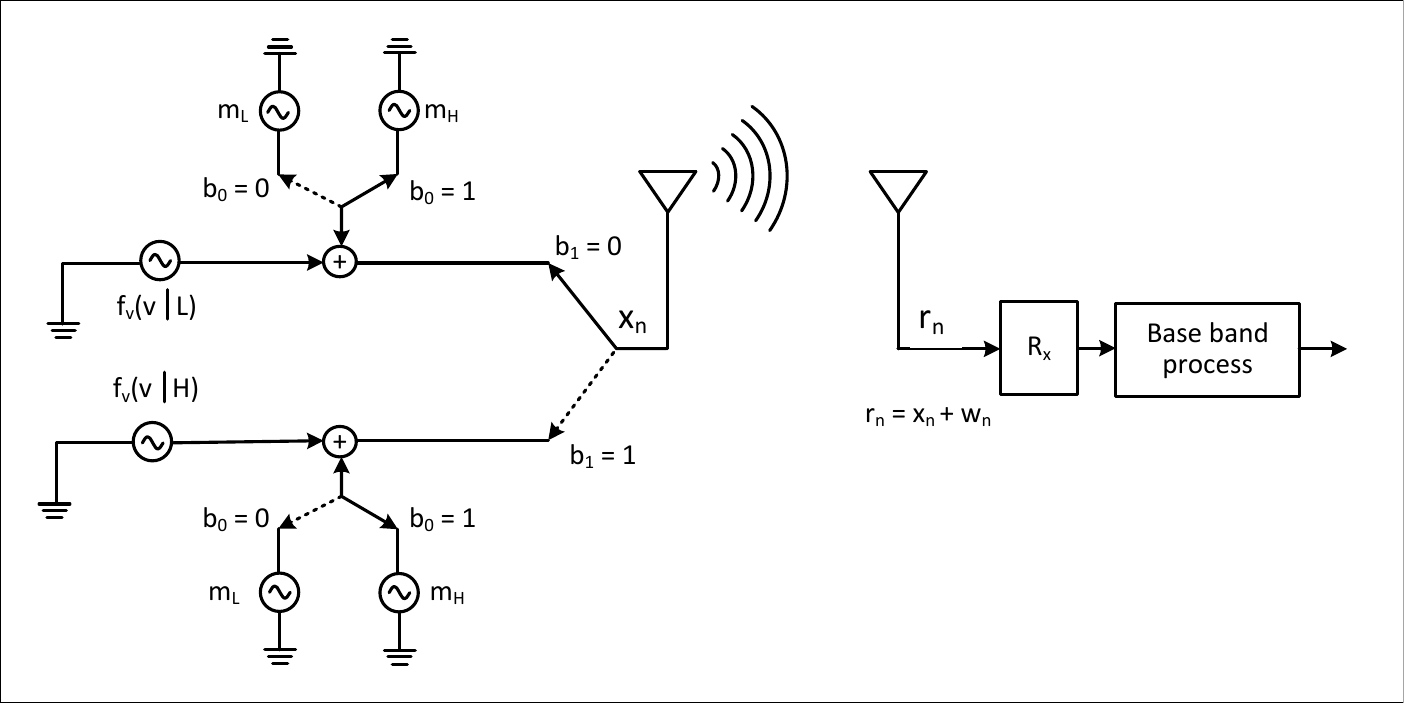}
\end{center}
\vspace{-0.5 cm}
\caption{Block diagram of generalized noise modulation scheme.}
\label{fig1}
\vspace{-0.5 cm}
\end{figure}
\begin{figure}
\begin{center}
\includegraphics[scale=0.38]{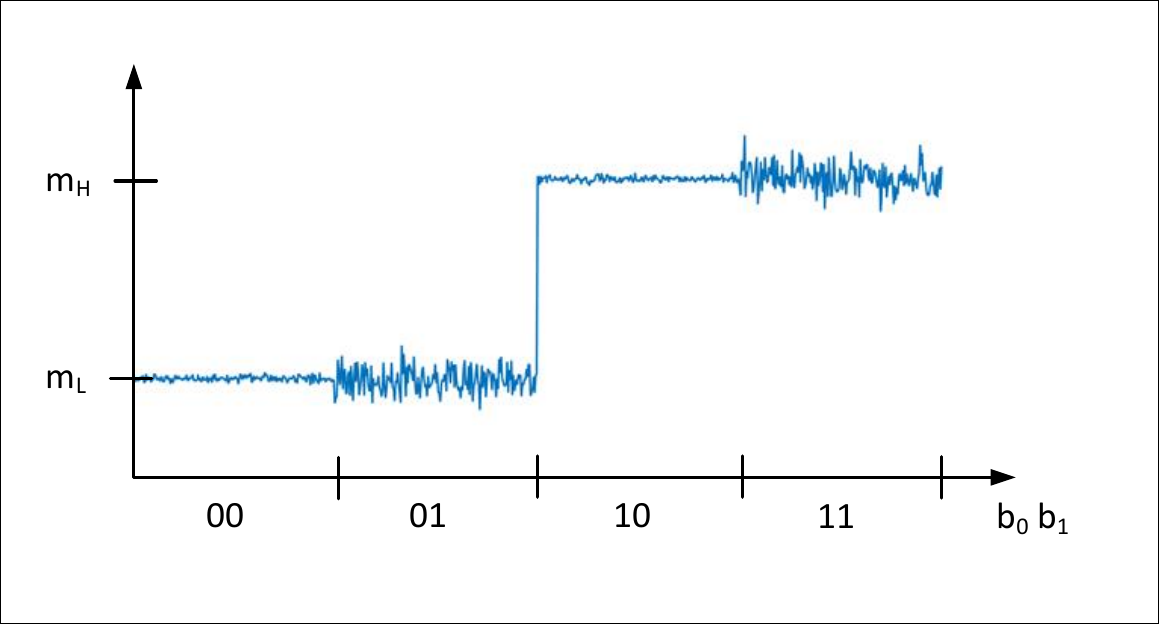}
\end{center}
\vspace{-0.5 cm}
\caption{Examples of waveforms in the impulsive MoTG case.}
\label{fig2}
\vspace{-0.7cm}
\end{figure}


In this letter, for the sake of tractability in derivations, we explore and simulate three specific cases: GG, GMoTG, and Laplacian. We provide a detailed design for all three cases, but derive the BEPs only for the GG and GMoTG cases, as the BEP calculations for the Laplacian case are challenging and are left for future work. Next, we present the QNM scheme for some distributions.


\subsection{GQNM with Gaussian distributions}\label{subsec:gg}
In this case, we have $f_v(v|L)=\frac{1}{\sigma_0\sqrt{2\pi}}\exp(-\frac{v^2}{2\sigma^2_0})$ and $f_v(v|H)=\frac{1}{\sigma_1\sqrt{2\pi}}\exp(-\frac{v^2}{2\sigma^2_1})$. This is the classical noise modulation case, where random thermal noise is generated by low- and high-resistors. We have $\sigma^2_0=4KTR_L\Delta f$ and $\sigma^2_1=4KTR_H\Delta f$, where $k=1.38\times 10^{-23}$ is the Boltzmann's constant, $T$ is the equivalent temperature in Kelvin, and $\Delta f$ is the bandwidth. In this case, the received sample $r_n$ is a Gaussian distribution $r_n\sim N(m_r,\sigma^2_r)$ with mean, $m_r$, and variance, $\sigma^2_r=\sigma^2_x+\sigma^2_w$, which are equal to
\begin{align}
\label{eq: mr}
m_r=\left\{
      \begin{array}{ll}
        m_L, & b_0b_1=00, \\
        m_L, & b_0b_1=01, \\
        m_H, & b_0b_1=10, \\
        m_H, & b_0b_1=11,
      \end{array}
    \right.
\end{align}
\begin{align}
\sigma^2_r=\left\{
      \begin{array}{ll}
        \sigma^2_w(1+\delta), & b_0b_1=00, \\
        \sigma^2_w(1+\alpha\delta), & b_0b_1=01, \\
        \sigma^2_w(1+\delta), & b_0b_1=10, \\
        \sigma^2_w(1+\alpha\delta), & b_0b_1=11,
      \end{array}
    \right.
\end{align}
where $\delta\triangleq\frac{\sigma^2_0}{\sigma^2_w}$, and $\alpha=\frac{R_H}{R_L}=\frac{\sigma^2_1}{\sigma^2_0}$. In the detector, we use the sample mean, $\hat{m}_r$, and sample variance, $\hat{\sigma}^2_r$, which are given by
\begin{align}
\label{eq: mg}
\hat{m}=\frac{1}{N}\sum_{n=1}^Nr_n\sim N(m_r,\frac{1}{N}\sigma^2_r),
\end{align}
\begin{align}
\label{eq: mv}
\hat{\sigma}^2_r=\frac{1}{N}\sum_{n=1}^Nr^2_n\sim N(m_{\sigma^2},\mathrm{Var}_{\sigma^2}),
\end{align}
where
\begin{align}
\label{eq: vv}
m_{\sigma^2}=\mathrm{E}\{\hat{\sigma}^2_r\}=\mathrm{E}\{r^2_n\}=m^2_r+\sigma^2_r,\\
\mathrm{Var}_{\sigma^2}=\mathrm{Var}(\hat{\sigma}^2_r)=\frac{1}{N}\mathrm{Var}(r^2_n),
\end{align}
in which $\mathrm{Var}(r^2_n)=m_{2,X}-m^2_{1,X}$, where $m_{k,X}=\mathrm{E}\{X^k\}$ is the $k$-th order moment of random variable $X$ with $X=r^2_n$. Then, we have $m_{1,X}=\mathrm{E}\{r^2_n\}=\mathrm{Var}(r_n)+m^2_r=\sigma^2_r+m^2_r$, and $m_{2,X}=\mathrm{E}\{r^4_n\}=m_{4,r_n}$. We know that if the random variable, $Z=r_n$, is Gaussian with $Z\sim N(\mu,\sigma^2)$, then we have $m_{4,Z}=\mu^4+6\mu^2\sigma^2+3\sigma^4$. Therefore, we have $m_{2,X}=m_{4,Z}=m^4_r+6m^2_r\sigma^2_r+3\sigma^4_r$. From (\ref{eq: vv}) and the previous calculations, we have
\begin{align}
\sigma^2_{var}&\triangleq\mathrm{Var}_{\sigma^2}=\frac{1}{N}\Big(m^4_r+6m^2_r\sigma^2_r+3\sigma^4_r-(m^2_r+\sigma^2_r)^2\Big)\nonumber\\
&=\frac{1}{N}\Big(4m^2_r\sigma^2_r+2\sigma^4_r\Big).
\end{align}

Following (\ref{eq: b1det}) and (\ref{eq: b0det}), we use the following detectors for $b_1$ and $b_0$
\begin{align}
\label{eq: det1}
\hat{b}_1=\left\{
            \begin{array}{ll}
              0, & \hat{\sigma}^2_r<\mathrm{Th}_v, \\
              1, & \hat{\sigma}^2_r>\mathrm{Th}_v,
            \end{array}
          \right.
\end{align}
and
\begin{align}
\label{eq: det2}
\hat{b}_0=\left\{
            \begin{array}{ll}
              0, & \hat{m}<\mathrm{Th}_m, \\
              1, & \hat{m}>\mathrm{Th}_m,
            \end{array}
          \right.
\end{align}
where $\mathrm{Th}_m=\frac{m_L+m_H}{2}$ is as general case, and $\mathrm{Th}_v=\frac{\sigma^2_L+\sigma^2_H}{2}$. The BEP, $\mathrm{p}_b$, is given by
\begin{align}
\label{eq: pbt}
\mathrm{p}_b=\frac{1}{2}(\mathrm{p}_{b,0}+\mathrm{p}_{b,1}),
\end{align}
where $\mathrm{p}_{b,l}=\mathrm{p}\{b_l\neq \hat{b}_l\}$ for $l=0,1$. Then, we have
\begin{align}
\label{eq: pbt1}
\mathrm{p}_{b,0}&=\mathrm{p}\{(b_0,b_1)=00\}\mathrm{p}\{error|(b_0,b_1)=00\}\nonumber\\
&+\mathrm{p}\{(b_0,b_1)=10\}\mathrm{p}\{error|(b_0,b_1)=10\}\nonumber\\
&+\mathrm{p}\{(b_0,b_1)=01\}\mathrm{p}\{error|(b_0,b_1)=01\}\nonumber\\
&+\mathrm{p}\{(b_0,b_1)=11\}\mathrm{p}\{error|(b_0,b_1)=11\},
\end{align}
where error event is $\hat{b}_0\neq b_0$. Since $\mathrm{p}\{(b_i,b_j)=\frac{1}{4}\}$ and with some manipulations, we have
\begin{align}
\label{eq: e1}
\mathrm{p}_{b,0}&=\frac{1}{4}\mathrm{p}\{\hat{m}>\mathrm{Th}_{m}|00\}+\frac{1}{4}\mathrm{p}\{\hat{m}<\mathrm{Th}_m|10\}\nonumber\\
&+\frac{1}{4}\mathrm{p}\{\hat{m}>\mathrm{Th}_m|01\}+\frac{1}{4}\mathrm{p}\{\hat{m}<\mathrm{Th}_m|11\}.
\end{align}

Since $\hat{m}$ is Gaussian as indicated in (\ref{eq: mg}), then $\sigma^2_r|_{00}=\sigma^2_r|_{10}=\sigma^2_w(1+\delta)$, and $\sigma^2_r|_{01}=\sigma^2_r|_{11}=\sigma^2_w(1+\alpha\delta)$. Thus, from (\ref{eq: e1}) and with some manipulations, we have
\begin{align}
\mathrm{p}_{b,0}&=\frac{1}{4}\Big[1-\mathrm{Q}\Big(\frac{m_L-\mathrm{Th}_m}{\sigma_w\sqrt{\frac{\delta+1}{N}}}\Big)\Big]+\frac{1}{4}\mathrm{Q}\Big(\frac{m_H-\mathrm{Th}_m}{\sigma_w\sqrt{\frac{1+\alpha\delta}{N}}}\Big)\nonumber\\
&+\frac{1}{4}\Big[1-\mathrm{Q}\Big(\frac{m_L-\mathrm{Th}_m}{\sigma_w\sqrt{\frac{\delta+1}{N}}}\Big)\Big]+\frac{1}{4}\mathrm{Q}\Big(\frac{m_H-\mathrm{Th}_m}{\sigma_w\sqrt{\frac{1+\alpha\delta}{N}}}\Big)\nonumber\\
&=\frac{1}{2}\Big[1-\mathrm{Q}\Big(\frac{m_L-\mathrm{Th}_m}{\sigma_w\sqrt{\frac{\delta+1}{N}}}\Big)+\mathrm{Q}\Big(\frac{m_H-\mathrm{Th}_m}{\sigma_w\sqrt{\frac{1+\alpha\delta}{N}}}\Big)\Big].
\end{align}

By following similar derivations for calculating the BEP of $b_0$, we have that the BEP of $b_1$ is defined as
\begin{align}
\label{eq: e2}
\mathrm{p}_{b,1}&=\frac{1}{4}\mathrm{p}\{\hat{\sigma}^2_r>\mathrm{Th}_{v}|00\}+\frac{1}{4}\mathrm{p}\{\hat{\sigma}^2_r<\mathrm{Th}_v|01\}\nonumber\\
&+\frac{1}{4}\mathrm{p}\{\hat{\sigma}^2_r>\mathrm{Th}_v|10\}+\frac{1}{4}\mathrm{p}\{\hat{\sigma}^2_r<\mathrm{Th}_v|11\}.
\end{align}

Since $\hat{\sigma}^2_r$ is Gaussian as indicated in (\ref{eq: mv}), $m_{\sigma^2}|_{00}=(m^2_r+\sigma^2_r)|{00}$, $m_{\sigma^2}|_{01}=(m^2_r+\sigma^2_r)|{01}$, $m_{\sigma^2}|_{10}=(m^2_r+\sigma^2_r)|{10}$, $m_{\sigma^2}|_{11}=(m^2_r+\sigma^2_r)|{11}$,   $\sigma^2_{var}|_{00}=\frac{1}{N}\Big(4m^2_r\sigma^2_r+2\sigma^4_r\Big)|_{00}$, $\sigma^2_{var}|_{01}=\frac{1}{N}\Big(4m^2_r\sigma^2_r+2\sigma^4_r\Big)|_{01}$, $\sigma^2_{var}|_{10}=\frac{1}{N}\Big(4m^2_r\sigma^2_r+2\sigma^4_r\Big)|_{10}$, and $\sigma^2_{var}|_{11}=\frac{1}{N}\Big(4m^2_r\sigma^2_r+2\sigma^4_r\Big)|_{11}$, from (\ref{eq: e2}) and with some manipulations, we have
\begin{align}
\mathrm{p}_{b,1}&=\frac{1}{4}\Big[1-\mathrm{Q}\Big(\frac{m_{\sigma^2}|_{00}-\mathrm{Th}_v}{\sigma_{var}|_{00}}\Big)\Big]+\frac{1}{4}\mathrm{Q}\Big(\frac{m_{\sigma^2}|_{01}-\mathrm{Th}_v}{\sigma_{var}|_{01}}\Big)\nonumber\\
&+\frac{1}{4}\Big[1-\mathrm{Q}\Big(\frac{m_{\sigma^2}|_{10}-\mathrm{Th}_v}{\sigma_{var}|_{10}}\Big)\Big]+\frac{1}{4}\mathrm{Q}\Big(\frac{m_{\sigma^2}|_{11}-\mathrm{Th}_v}{\sigma_{var}|_{11}}\Big).
\end{align}

\subsection{GQNM with MoTG distributions}
Here, we have $f_v(v|L)=pN(0,\sigma^2_{0L})+(1-p)N(0,\sigma^2_{1L})$ and $f_v(v|H)=pN(0,\sigma^2_{0H})+(1-p)N(0,\sigma^2_{1H})$, where we assume $\sigma_{0L}<\sigma_{1L}$, $\sigma_{0H}<\sigma_{1H}$, $\sigma_{0L}<\sigma_{0H}$, and $\sigma_{1L}<\sigma_{1H}$. In this case, the value of $m_r$ is the same as in (\ref{eq: mr}), and the value of $\sigma^2_r=\sigma^2_w+\sigma^2_{x_n}$ is
\begin{align}
\sigma^2_{r,MoG}=\left\{
      \begin{array}{ll}
        \sigma^2_w+p\sigma^2_{0L}+(1-p)\sigma^2_{1L}, & b_0b_1=00, \\
        \sigma^2_w+p\sigma^2_{0H}+(1-p)\sigma^2_{1H}, & b_0b_1=01, \\
        \sigma^2_w+p\sigma^2_{0L}+(1-p)\sigma^2_{1L}, & b_0b_1=10, \\
        \sigma^2_w+p\sigma^2_{0H}+(1-p)\sigma^2_{1H}, & b_0b_1=11.
      \end{array}
    \right.
\end{align}

Moreover, based on the Central Limit Theorem (CLT), for a large $N$, we have
\begin{align}
\hat{m}\sim N(m_r,\sigma^2_{r,MoG}), \hat{\sigma}^2_r\sim N(m_{\sigma^2_{MoG}},\frac{1}{N}\mathrm{Var}(r^2_n)),
\end{align}
where $m_{\sigma^2_{MoG}}=\mathrm{E}\{r^2_n\}=\sigma^2_{r,MoG}+m^2_r$ and
\begin{align}
\mathrm{Var}(r^2_n)=\mathrm{E}\{r^4_n\}-\mathrm{E}^2\{{r^2_n}\}=\mathrm{E}\{r^4_n\}-(\sigma^2_{r,MoG}+m^2_r)^2,
\end{align}
in which
\begin{align}
\mathrm{E}\{r^4_n\}=\mathrm{E}\{(x_n+w_n)^4\}=\mathrm{E}\{x^4_n\}+6\mathrm{E}\{x^2_n\}\sigma^2_w+3\sigma^4_w,
\end{align}
with $\mathrm{E}\{x^2_n\}=\sigma^2_{r,MoG}+\sigma^2_w$, which results in
\begin{align}
\mathrm{E}\{x^4_n\}=\mathrm{E}\{(Y+m_r)^4\}=\mathrm{E}\{\mathrm{Y^4}\}+6m^2_r\mathrm{E}\{Y^2\}+m^4_r,
\end{align}
where $Y=x_n-m_r$ is the MOTG variable. We have $\mathrm{E}\{Y^2\}=\mathrm{E}\{x^2_n\}-2m_r\mathrm{E}\{x_n\}+m^2_r$. Since $\mathrm{E}\{x_n\}=m_r$, then $\mathrm{E}\{Y^2\}=\mathrm{E}\{x^2_n\}-m^2_r=\mathrm{Var}(x_n)+\mathrm{E}^2\{x_n\}-m^2_r=\mathrm{Var}(x_n)$. This way, we have
\begin{align}
\mathrm{E}\{Y^2\}=\left\{
                      \begin{array}{ll}
                        p\sigma^2_{0L}+(1-p)\sigma^2_{1L}, & b_0b_1=00, \\
                        p\sigma^2_{0H}+(1-p)\sigma^2_{1H}, & b_0b_1=01, \\
                        p\sigma^2_{0L}+(1-p)\sigma^2_{1L}, & b_0b_1=10, \\
                        p\sigma^2_{0H}+(1-p)\sigma^2_{1H}, & b_0b_1=11,
                      \end{array}
                    \right.
\end{align}
and
\begin{align}
\mathrm{E}\{Y^4\}=\mathrm{E}\{v^4_n\}=\left\{
                      \begin{array}{ll}
                        3(\sigma^4_{0L}+\sigma^4_{1L}), & b_0b_1=00, \\
                        3(\sigma^4_{0H}+\sigma^4_{1H}), & b_0b_1=01, \\
                        3(\sigma^4_{0L}+\sigma^4_{1L}), & b_0b_1=10, \\
                        3(\sigma^4_{0H}+\sigma^4_{1H}), & b_0b_1=11.
                      \end{array}
                    \right.
\end{align}

The detectors are the same as in (\ref{eq: det1}) and (\ref{eq: det2}), with a mean threshold that is the same as before, but with a different variance threshold equal to
\begin{align}
\mathrm{Th}_v=\frac{(p\sigma^2_{0L}+(1-p)\sigma^2_{1L})+(p\sigma^2_{0H}+(1-p)\sigma^2_{1H})}{2}.
\end{align}

To calculate the BEP, we follow the same rationale as in \ref{subsec:gg}. Thus, after applying similar calculations as before, we find
\begin{align}
\mathrm{p}_{b,0,MoG}&=\frac{1}{2}\Big[1-\mathrm{Q}\Big(\frac{m_L-\mathrm{Th}_m}{\sigma_{r,MoG}|_{00}}\Big)+\mathrm{Q}\Big(\frac{m_H-\mathrm{Th}_m}{\sigma_{r,MoG}|_{01}}\Big)\Big],
\end{align}
and
\begin{align}
&\mathrm{p}_{b,1,MoG}=\nonumber\\
&\frac{1}{4}\Big[1-\mathrm{Q}\Big(\frac{m_{\sigma^2_{MoG}}|_{00}-\mathrm{Th}_v}{\sqrt{\frac{\mathrm{Var}(r^2_n)|_{00}}{N}}}\Big)\Big]+\frac{1}{4}\mathrm{Q}\Big(\frac{m_{\sigma^2_{MoG}}|_{01}-\mathrm{Th}_v}{\sqrt{\frac{\mathrm{Var}(r^2_n)|_{01}}{N}}}\Big)\nonumber\\
&+\frac{1}{4}\Big[1-\mathrm{Q}\Big(\frac{m_{\sigma^2_{MoG}}|_{10}-\mathrm{Th}_v}{\sqrt{\frac{\mathrm{Var}(r^2_n)|_{10}}{N}}}\Big)\Big]+\frac{1}{4}\mathrm{Q}\Big(\frac{m_{\sigma^2_{MoG}}|_{11}-\mathrm{Th}_v}{\sqrt{\frac{\mathrm{Var}(r^2_n)|_{11}}{N}}}\Big).
\end{align}

\subsection{GQNM with Laplace distributions}
Here, we have $f_v(v|L)=\frac{1}{2\lambda_0}\exp(-\frac{|v|}{2\lambda^2_0})$ and $f_v(v|H)=\frac{1}{2\lambda_1}\exp(-\frac{|v|}{2\lambda^2_1})$ with the assumption of $\lambda_0<\lambda_1$. In this case, the received sample, $r_n$, follows an unknown distribution, $r_n\sim f(r_n)$, with mean $m_{r,L}=m_r$ indicated in (\ref{eq: mr}) and variance $\sigma^2_{r,L}=\sigma^2_{x,L}+\sigma^2_w$, which is defined as
\begin{align}
\sigma^2_{r,L}=\left\{
      \begin{array}{ll}
        \sigma^2_w(1+\delta_{Lap}), & b_0b_1=00, \\
        \sigma^2_w(1+\alpha_{Lap}\delta_{Lap}), & b_0b_1=01, \\
        \sigma^2_w(1+\delta_{Lap}), & b_0b_1=10, \\
        \sigma^2_w(1+\alpha_{Lap}\delta_{Lap}), & b_0b_1=11,
      \end{array}
    \right.
\end{align}
where $\delta_{Lap}\triangleq\frac{\sigma^2_{0,Lap}}{\sigma^2_w}=\frac{2\lambda^2_0}{\sigma^2_w}$, and $\alpha_{Lap}=\frac{\sigma^2_{1,L}}{\sigma^2_{0,L}}=\frac{\lambda^2_1}{\lambda^2_0}$. The detectors are the same as before, and we omit them for brevity. The only difference is the variance threshold, which is $\mathrm{Th}_v=\lambda^2_0+\lambda^2_1$. Calculating the BEP in this case is challenging and is left for future work. However, we include this subsection as it is necessary for the simulation of GLAP in the experiments.


\textbf{Remark 1:} To fairly compare the three noise modulators with GG, GMoTG, and GLAP, we calculate the power of the transmitter and take the parameters of the systems to have the same transmit power. Therefore, the transmit power of GG is equal to $P_{GG}=\mathrm{E}\{x^2_n\}=\frac{1}{4}\Big[\mathrm{E}\{x^2_n|00\}+\mathrm{E}\{x^2_n|01\}+\mathrm{E}\{x^2_n|10\}+\mathrm{E}\{x^2_n|11\}\Big]$ which is simplified to $P_{GG}=\frac{1}{2}(m^2_L+m^2_H+\sigma^2_L+\sigma^2_H)$. The transmit powers of GMoTG and GLAP are equal to
\begin{align}
\label{eq: PMOG}
&P_{GMoTG}=\mathrm{E}\{x^2_n\}\nonumber\\
&=\frac{1}{4}\Big[\mathrm{E}\{x^2_n|00\}+\mathrm{E}\{x^2_n|01\}+\mathrm{E}\{x^2_n|10\}+\mathrm{E}\{x^2_n|11\}\Big]\nonumber\\
&=\frac{1}{4}\Big[(p(m^2_L+\sigma^2_{0L})+(1-p)(m^2_L+\sigma^2_{0H}))\nonumber\\
&+(p(m^2_L+\sigma^2_{1L})+(1-p)(m^2_L+\sigma^2_{1H}))\nonumber\\
&+(p(m^2_H+\sigma^2_{0L})+(1-p)(m^2_H+\sigma^2_{0H}))\nonumber\\
&+(p(m^2_H+\sigma^2_{1L})+(1-p)(m^2_H+\sigma^2_{1H}))\Big],
\end{align}
and
\begin{align}
\label{eq: PLAP}
&P_{GLAP}=\mathrm{E}\{x^2_n\}=\nonumber\\
&\frac{1}{4}\Big[\mathrm{E}\{x^2_n|00\}+\mathrm{E}\{x^2_n|01\}+\mathrm{E}\{x^2_n|10\}+\mathrm{E}\{x^2_n|11\}\Big]=\nonumber\\
&\frac{1}{4}\Big[(m^2_L+2\lambda^2_0)+(m^2_L+2\lambda^2_1)+(m^2_H+2\lambda^2_0)+(m^2_H+2\lambda^2_1)\Big]\nonumber\\
&=\frac{1}{2}(m^2_L+m^2_H+2\lambda^2_0+2\lambda^2_1).
\end{align}

\section{Simulation Results}
\label{sec: Simulation}
In this section, the performance of the proposed GQNM scheme is investigated. We simulated the three versions of GQNM: GG, GMoTG, and GLAP. Their parameters are set to have distinguishability of distributions. It is provided by trial and error, as presented next. In all schemes, we use $m_L=1e-3$ and $m_H=1e-2$. The parameters of GG are $\sigma_{0}=1e-3$ and $\sigma_1=20e-3$. The parameters of GMoTG are $\sigma_{0L}=5e-4$, $\sigma_{1L}=1e-3$, $\sigma_{0H}=5e-3$. The fourth parameter $\sigma_{1H}$ is selected as $\sigma_{1H}=21e-3$ from (\ref{eq: PMOG}) to the MoTG case, which has the same transmitted power as the GG case. The parameters of GLAP are $\lambda_0=1e-4$ and $\lambda_1$ is selected as $\lambda_1= 14.2e-3$ from (\ref{eq: PLAP}), so that the Laplacian case has the same transmit power as the other schemes.
For a fair comparison, we set the average transmit power to be equal for all three schemes, ensuring they are compared at the same signal-to-noise ratio (SNR).
The thresholds are set as defined in Section \ref{sec: prop}. Unless otherwise stated, the noise standard deviation is $\sigma_w=2\times10^{-5}$ and the number of samples per bit duration is $N=10$. The performance is evaluated using the general and individual bit error probabilities (BEPs): $\mathrm{p}b$, $\mathrm{p}{b,0}$, and $\mathrm{p}_{b,1}$.

\begin{figure}
\begin{center}
\includegraphics[width=7cm]{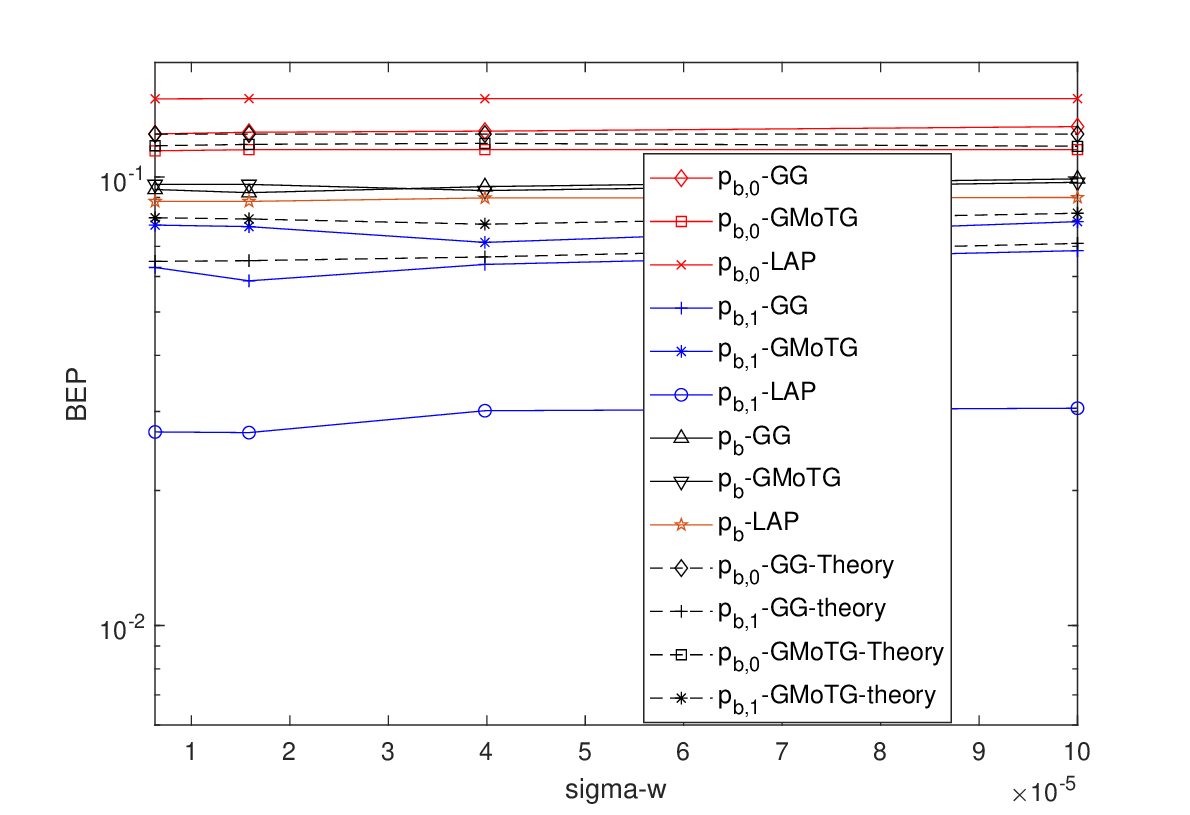}
\end{center}
\vspace{-0.5 cm}
\caption{Simulated and theoretical BEPs versus noise standard deviation, $\sigma_w$.}
\label{fig4}
\vspace{-0.5 cm}
\end{figure}

Two experiments were performed. In the first one, depicted in Fig.~\ref{fig4}, the effect of channel noise, $w_n$, on simulated and theoretical (dashed lines) BEPs is evaluated. We vary the $\sigma_n$ from $10^{-5.2}\approx 6.63e-6$ to $1e-4$.
It can be seen that GLAP has the lowest $\mathrm{p}_{b,1}$ while GMoTG has the lowest $\mathrm{p}_{b,0}$. However, the total bit error probability is approximately the same for the three schemes, with GLAP having the lowest $\mathrm{p}_b$. Additionally, the figure shows that the standard deviation $\sigma_w$ does not affect the performance of the three schemes, i.e., all three schemes are independent of $\sigma_w$. This result also demonstrates a good superiority of GLAP in comparison to the other schemes regarding $\mathrm{p}_{b,1}$, which shows that the generalized Laplacian scheme can be used to decrease the BEP in one sub-bit of a symbol composed of two bits. If we have two sub-bit channels of data, the more important data channel\footnote{Importance is from the viewpoint of a specific data communication application. For example, in satellite communication, the control/command data may be more important than telemetry data.} can be modulated in this sub-bit, which is in the variance dimension. GLAP performs better in the $\mathrm{p}_{b,1}$ dimension because the heavy-tailed nature of the Laplacian distribution causes its sample variance estimator to be more "decisive" when in the high-variance state.

\begin{figure}
\begin{center}
\includegraphics[width=7cm]{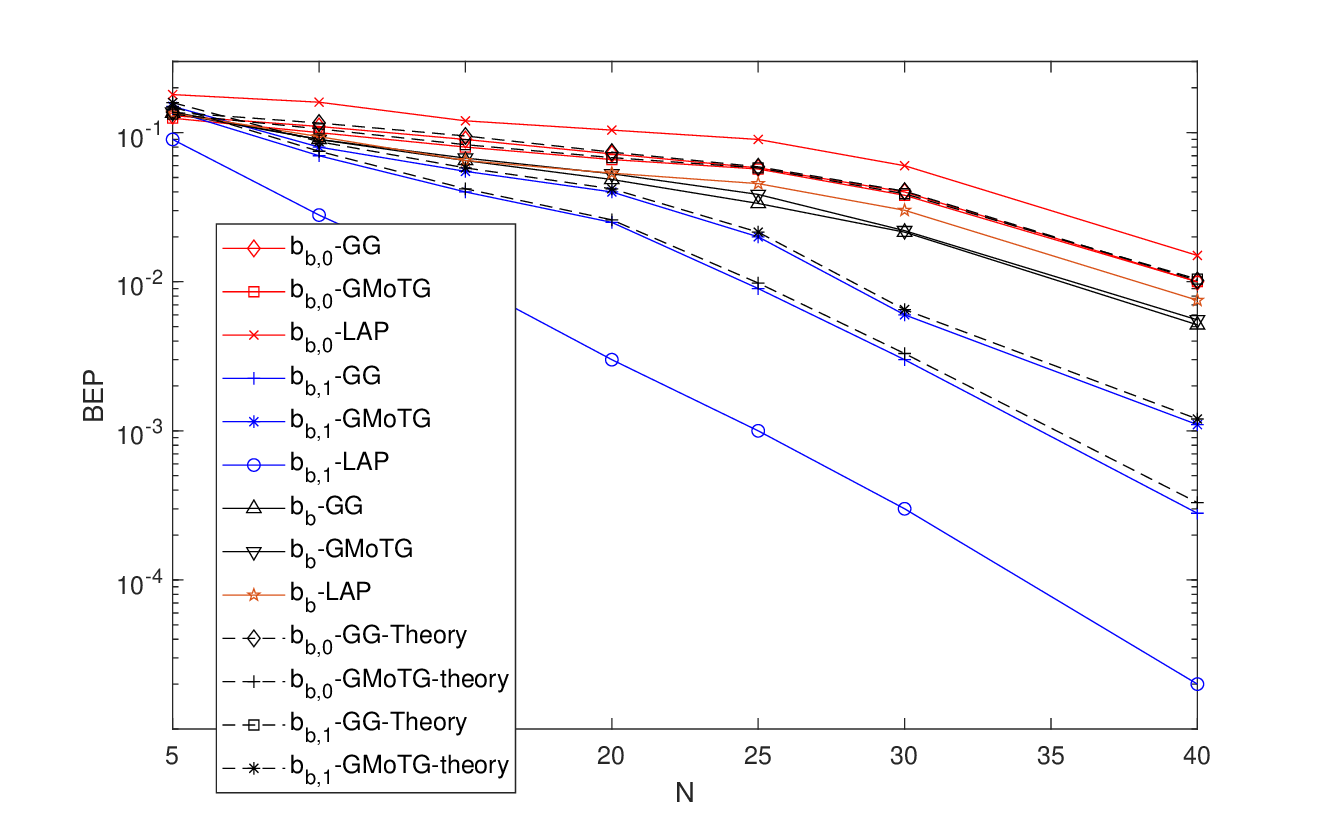}
\end{center}
\vspace{-0.5 cm}
\caption{Simulated and theoretical BEPs versus $N$.}
\label{fig5}
\vspace{-0.5 cm}
\end{figure}

The second experiment, shown in Fig.~\ref{fig5}, compares the simulated and theoretical (dashed lines) BEPs versus $N$. It shows the effect of $N$ on the performance of the schemes. The parameter $N$ is varied between 5 and 40. The results confirm almost similar results to the first experiment. It shows that by increasing $N$, the BEPs decrease. It also demonstrates a good match between theoretical and simulated BEPs.

%

\section{Conclusion and future work}
\label{sec: con}

This letter generalized the concept of binary noise modulation into a generalized quadratic noise modulation (GQNM) scheme. The proposed framework introduces two key novelties: first, the use of two distinct voltage biases to encode an additional bit in the signal's mean, and second, the utilization of synthetic non-Gaussian noise distributions, such as the MoG and Laplacian distributions, in addition to the conventional Gaussian noise. This approach effectively doubles the data rate of traditional noise modulation by enabling simultaneous discrimination in both the mean and variance dimensions. For the specific study cases of GG and GMoTG noise, closed-form expressions for the BEP were rigorously derived, providing a solid theoretical foundation for performance analysis. Simulation results demonstrated the advantages of the proposed GQNM scheme. Furthermore, they indicated that employing non-Gaussian distributions, such as the Laplacian and MoG, can offer performance benefits alongside the inherent data rate doubling. While the BEP for the Laplacian case was identified as a challenge for future work, this study opens the door to exploring even higher-order noise modulations that leverage additional distinguishable parameters of complex distributions for further gains in data rate and reliability. 



\begin{thebibliography}{1}


\bibitem{Kish05}
L. B. Kish,
\newblock ``Stealth communication: Zero-power classical communication, zero-quantum quantum communication and environmental-noise communication,''
\newblock {\em Apply. Phys. Letter}, vol. 87, no. 23, Dec. 2005.

\bibitem{Kish06}
L. B. Kish,
\newblock ``Totally secure classical communication utilizing Johnson(-like) noise and Kirchoff's law,''
\newblock {\em Phys. Letter A}, vol. 352, pp. 178--182, Sep. 2006.

\bibitem{Kish06_1}
L. B. Kish, et al,
\newblock ``Totally secure classical networks with multipoint telecloning (teleportation) of classical bits through loops with Johnson-like noise,''
\newblock {\em Fluct. Noise Lett.}, vol. 6, no. 2, pp. C9--C21, Apr. 2006.

\bibitem{Ming08}
R. Mingesz, et al
\newblock ``Johnson(-like) noise Kirchhoff loop-based secure classical communicator characteristics, for ranges of two to two thousand kilometers, via model-line,''
\newblock {\em Phys. Letter A}, vol. 372, no. 7, pp. 978--984, Feb. 2008.


\bibitem{Kish10}
L. B. Kish, et al,
\newblock ``Noise in the wire: The real impact of wire resistance for the Johnson(-like) noise-based secure communicator,''
\newblock {\em Phys. Letter A.}, vol. 374, no. 21, pp. 2140--2142, Apr. 2010.

\bibitem{Saez13}
Y. Saez, and L. B. Kish,
\newblock ``Errors and their mitigation at the Kirchhoff law-Johnson-noise secure key exchange,''
\newblock {\em PLOS ONE}, vol. 8, no. 11, pp. 1--7, Nov. 2013.

\bibitem{Saez13_1}
Y. Saez, et al,
\newblock ``Current and voltage based bit errors and their combined mitigation for the Kirchhoff-law Johnson-noise secure key exchange,''
\newblock {\em J. Comput. Electron.}, vol. 13, no. 1, pp. 271--277, Mar. 2013.



\bibitem{Ging14}
Z. Gingl, and R. Mingesz,
\newblock ``Noise properties in the ideal Kirchhoff-law Johnson-noise secure communication system''
\newblock {\em PLOS ONE}, vol. 9, no. 4, Apr. 2014.

\bibitem{Vadai15}
G. Vadai, et al,
\newblock ``Generalized Kirchhoff-law Johnson-noise (KLJN) secure key exchange system using arbitrary resistors''
\newblock {\em Sci.Rep}, vol. 5, Sep. 2015.

\bibitem{Ferd20}
S. Ferdous, et al,
\newblock ``Comments on the Generalized KJLN key exchanger with arbitrary resistors: Power, impedance, security,''
\newblock {\em Fluct. Noise Lett.}, vol. 20, no. 1, Nov. 2020.


\bibitem{Kape22}
Z. Kapetanovic, et al,
\newblock ``Communication by means of modulated Johnson noise''
\newblock {\em PNAS (Computer sciences)}, vol. 119, no. 49, 2022.

\bibitem{Basar23}
E. Basar,
\newblock ``Communication by Means of Thermal Noise: Towards Networks with Extremely Low Power Consumption,''
\newblock {\em IEEE Trans. on Communication}, vol. 71, no. 2, Feb. 2023.


\bibitem{Tasci25}
R. A. Tasci, et al
\newblock ``Flip-KLJN: Random Resistance Flipping for Noise-Driven Secure Communication,''
\newblock {\em IEEE Trans. on Communication}, Early Access, July 2025.

\bibitem{Salem25}
M. A. Salem, et al
\newblock ``A KLJN-Based Thermal Noise Modulation Scheme With Enhanced Reliability for Low-Power IoT Communication,''
\newblock {\em IEEE Open Journal of the Communications Society}, vol. 6, August 2025.


\bibitem{Basar24}
E. Basar,
\newblock ``Noise modulation,''
\newblock {\em IEEE Wireless Communication Letters}, vol. 13, no. 3, March 2024.

\bibitem{Alshaw24}
M. K. Alshawaqfeh, et al,
\newblock ``Thermal Noise Modulation: Optimal Detection and Performance Analysis,''
\newblock {\em IEEE Communications Letters}, vol. 28, no. 12, Dec. 2024.

\bibitem{Anjos25}
A. A. dos Anjos, and H. S. Silva,
\newblock ``On-Off Digital Noise Modulation,''
\newblock {\em IEEE Wireless Communication Letters}, Early Access, 2025.

\bibitem{Yapici25}
E. Yapici, et al,
\newblock ``Noise Modulation Over Wireless Energy Transfer: JEIH-NoiseMod,''
\newblock {\em IEEE Wireless Communication Letters}, Early Access, 2025.












%
%
%




\end{thebibliography}
\end{document}